\documentclass[journal,onecolumn]{IEEEtran}
\usepackage{amsmath,amsfonts,amssymb,mathtools,mathrsfs}
\usepackage{array}
\usepackage{hyperref}
\usepackage{svg}
\usepackage{textcomp}
\usepackage{stfloats}
\usepackage{url}
\usepackage{verbatim}
\usepackage{graphicx,color}
\usepackage{subfigure}
\usepackage{cite}
\usepackage[english]{babel}
\usepackage{amsthm}
\usepackage{algorithm,algpseudocode}

\newcommand{\e}{e}

\newcommand{\vct}[1]{\boldsymbol{#1}}
\newcommand{\mtx}[1]{\boldsymbol{#1}}

\renewcommand{\H}{\mathrm{H}}

\newcommand{\set}[1]{\mathcal{#1}}

\newcommand{\vb}{\vct{b}}

\newcommand{\ve}{\vct{e}}

\newcommand{\vn}{\vct{n}}

\newcommand{\vp}{\vct{p}}

\newcommand{\vr}{\vct{r}}
\newcommand{\vs}{\vct{s}}

\newcommand{\vw}{\vct{w}}
\newcommand{\vx}{\vct{x}}
\newcommand{\vy}{\vct{y}}

\newcommand{\valpha}{\vct{\alpha}}
\newcommand{\vbeta}{\vct{\beta}}

\newcommand{\mA}{\mtx{A}}

\newcommand{\mF}{\mtx{F}}
\newcommand{\mG}{\mtx{G}}
\newcommand{\mH}{\mtx{H}}

\newcommand{\mL}{\mtx{L}}

\newcommand{\mP}{\mtx{P}}
\newcommand{\mQ}{\mtx{Q}}

\newcommand{\mU}{\mtx{U}}

\newcommand{\mId}{{\bf I}}

\newcommand{\setI}{\set{I}}

\newcommand{\RN}[1]{%
  \textup{\uppercase\expandafter{\romannumeral#1}}%
}

\begin{document}

\title{A Fast Broadband Beamspace Transformation}

\author{Nakul Singh, Coleman DeLude, Mark A. Davenport, and Justin Romberg
\thanks{N. Singh, C. DeLude, M. Davenport, and J. Romberg, are all with the School of Electrical and Computer Engineering at the Georgia Institute of Technology.  Email: \{nsingh360, cdelude3, mdav, jrom@ece\}.gatech.edu. This work was supported by CogniSense, one of seven centers in JUMP 2.0, a Semiconductor Research Corporation (SRC) program sponsored by DARPA, and a grant from Lockheed Martin.}
}



\maketitle

\begin{abstract}

We present a new computationally efficient method for multi-beamforming in the broadband setting.  Our ``fast beamspace transformation'' forms $B$ beams from $M$ sensor outputs using a number of operations per sample that scales linearly (to within logarithmic factors) with $M$ when $B\sim M$.  While the narrowband version of this transformation can be performed efficiently with a spatial fast Fourier transform, the broadband setting requires coherent processing of multiple array snapshots simultaneously.

Our algorithm works by taking $N$ samples off of each of $M$ sensors and encoding the sensor outputs into a set of coefficients using a special non-uniformly spaced Fourier transform.  From these coefficients, each beam is formed by solving a small system of equations that has Toeplitz structure.  The total runtime complexity is $\mathcal{O}(M\log N+B\log N)$ operations per sample, exhibiting essentially the same scaling as in the narrowband case.
Our method offers significantly higher accuracy and lower computational complexity than delay-and-sum filtering techniques and significantly higher accuracy and lower latency than standard frequency-domain techniques based on sub-band processing.

An additional benefit of the proposed approach is that it allows for a careful error analysis; we give concrete formulas for the bias and variance of the beamformer given the array parameters.  We also provide a host of numerical experiments demonstrating the algorithm's favorable computational scaling and high accuracy. Finally, we demonstrate how tasks such as interpolating to ``off-grid" angles and nulling an interferer are more computationally efficient when performed directly in beamspace.

\end{abstract}


\section{Introduction}

Modern array processing applications have been consistently pushed towards operating with higher bandwidth, lower latency, and larger arrays. A key step at the center of a variety of tasks --- including direction of arrival estimation, beam training, interference nulling, and clutter mitigation --- is the essential act of simultaneous \emph{multi-beamforming}. This step spatially separates the signals received at the array by forming a large number of beams steered in distinct directions. 
In the narrowband regime, the fast Fourier transform (FFT) can be used to efficiently form many beams simultaneously on a fixed grid of angles: a spatial FFT is applied to each array snapshot independently, the result of which is a sample of a beam at each angle in the grid.
In the broadband regime, delays across sensors in the array can no longer be approximated by phase shifts, breaking the key approximation that allows us to use the spatial Fourier transform for beamforming.
In this paper we explore an alternative approach to the problem of efficient \emph{broadband} multi-beamforming.

Our proposed \emph{fast beamspace transformation} (FBST) offers a low-complexity alternative to direct calculation of multiple broadband beams. In particular, the algorithm scales log-linearly with the array size ($M$), samples ($N$), and number of beams ($B$); the per-sample runtime complexity is $\mathcal{O}(M\log N+B\log N)$.  
In contrast, per-sample direct beam computation scales with the product of these variables, $\mathcal{O}(MB)$. The FBST supports both linear and planar array configurations and seamlessly transitions between broadband and narrowband signal scenarios. As demonstrated in our numerical section using very large arrays (e.g., $>2^8$ elements), FBST provides a 100-fold speedup over delay and sum beamforming.

The FBST algorithm operates in two stages. In the first stage, a generalization of the DFT known as the chirp-$z$ transform is used to project the recorded array data into a ``beamspace," capturing information from multiple directions simultaneously. 
The end result is a set of Fourier-like coefficients for each beam that can now be processed independently.

In the second stage, we solve a system of equations to form each beam.  Through a careful choice of basis representation, this system has Toeplitz structure, making its inverse both fast to compute and (more importantly in our case) fast to apply.  Each beam requires a system solve, and the solutions can be computed in parallel.
%
%
The resulting signal reconstructions enjoy near-ideal (e.g., perfect true time delay) beamforming performance while also being drastically more efficient than direct multi-beamforming.

While the idea of efficiently mapping array measurements into beamspace has long been established in the context of narrowband beamforming \cite{vantrees2002optimum}, existing broadband beamspace transformation methods are either inefficient when scaling to large numbers of beams, inaccurate, or incur significant latency.   
Our framework has the same order per-sample complexity as the narrowband case, offers near-optimal accuracy across all SNR regimes, operates from small batches of samples (and thus incurs low latency), and can be seamlessly adapted to any signal bandwidth.
These characteristics make our beamspace framework well-suited for a wide range of array processing tasks. We illustrate this below using two particular applications: off-grid angle interpolation and interference cancellation. We note, however, that there are many other processing tasks that can be accelerated in beamspace.


The remainder of this paper is organized as follows. Section~\ref{sec:related} provides a thorough overview of existing fast multi-beamforming techniques. In Section~\ref{sec:form} we formulate the problem and provide the necessary models for understanding the FBST. Section~\ref{sec:transform} shows how to efficiently perform the beamspace transformation using chirp-$z$ based operations. This is followed by Section~\ref{sec:fi} which mathematically formulates the Toeplitz inverse problem required to recover signals from the beamspace. A host of numerical experiments and a theoretical error analysis are presented in sections \ref{sec:num_exp} and \ref{sec:err}, respectively.\footnote{See \href{https://github.com/nsingh360/FBST_Code}{https://github.com/nsingh360/Code-FBST} for the code used for these experiments.} In Section~\ref{sec:beam_apps}, we show how this technique can be leveraged in interference cancellation and off-grid angle interpolation. Finally, Section~\ref{sec:nufft_1} provides a straightforward way to extend the FBST framework to non-uniform linear and planar arrays. 

A preliminary version of this work appeared in \cite{fbst_conf}. This paper significantly extends the conference version in several aspects. Specifically, we include a detailed discussion on basis parameter selection, provide a rigorous error analysis of Fourier extensions, present new application results demonstrating computational speedups, and extend the proposed framework to non-uniform linear and planar arrays.


\section{Related Work}
\label{sec:related}

Low-complexity multi-beamforming for the narrowband regime uses specific projection matrices to form multiple beams in fixed directions \cite{10223668, Van1988BeamformingAV}. Since the steering vectors are just phase shifts, one obvious choice is to use DFT as a spatial transform over the array aperture for each time sample, where each DFT bin corresponds to a beam direction. 

To form multiple beams by applying the DFT matrix, the angles for the fixed beams are sampled such that the resultant grid is rectangular. Therefore, forming $B$ beams on a $B$ element array becomes the same as evaluating the $B$ point DFT of the incoming signal \cite{7942144}. A naive implementation of the DFT requires $\mathcal{O}(B^2)$ multiplications, whereas employing the Fast Fourier Transform (FFT) reduces the complexity to $\mathcal{O}(B\log B)$. While the FFT significantly reduces the number of multiplications, these operations still pose challenges for hardware implementations due to energy consumption and circuit area constraints \cite{Suarez2014MultiBeamRA}.

To reduce power consumption and circuit complexity, another alternative to FFT is to use an approximate DFT (ADFT) \cite{7112356, 9064580, 8542737, 8439422, 10132482} that maintains the same level of performance while avoiding explicit multiplication operations. The approach uses parametric optimization to derive an approximation to the Fourier matrix that can be applied to the array measurements using additions and bit shift operations. Another feasible alternative is the Discrete Cosine Transform (DCT) and its approximate versions \cite{article_dct,7169222}, which provide an energy-efficient means of performing multi-beamforming with lower computational overhead. DCT-based beamforming methods capitalize on the real-valued nature of the transformation, reducing complexity while maintaining comparable performance to DFT-based beamforming. The methods above provide viable approaches to low-complexity multi-beamforming for narrowband signals; however, they cannot be applied directly to broadband signals, as there is no straightforward analog to the steering vector in the broadband case. To remedy this, sub-band processing is a popular method of extending narrowband beamformers to broadband regimes\cite{Weib2009DigitalAntennas}. 

{In sub-band processing samples are buffered at each sensor, an FFT is applied to each buffer, and narrowband beamforming is performed across each frequency bin. After the frequency-domain operations, an inverse FFT is used to reconstruct a block of beamformed time samples. Achieving high frequency resolution and eliminating beam-squint demands collecting a large number of samples at each sensor. The exact number scales with the signal's bandwidth, and is often on the order of thousands for broadband use cases. Therefore this process carries substantial memory requirements in addition to inducing significant latency. This is apparent in the numerical experiments in Section~\ref{sec:num_exp}, which show that using a limited number of samples leads to considerable degradation in beamforming performance, even when the number of samples is in the thousands. Moreover, frequency-domain processing can create distortions across successive blocks of beamformed samples, complicating their application to real-time streaming data \cite{10835170}. This is in addition to distortions due to the finite extent of the signal, a property common to classic frequency domain approaches and their time domain analogs \cite{compton1988th}. In contrast, our method requires far fewer samples to ensure beam-squint-free performance and highly accurate signal reconstructions.}
%

Similar efforts have been directed toward speeding up delay and sum based broadband multi-beamforming. In \cite{pruned, quadtree}, a multi-stage recursive formulation is proposed that speeds up the delay and sum operation by coherently combining sensors in the spatial domain at each stage. An alternative strategy involves transitioning to the frequency domain, as described in the previous paragraph. In this context, \cite{vander1, vander2} uses a Vandermonde matrix to represent frequency-dependent phase delays and leverages a sparse factorization for faster processing. While these methods enable faster broadband multi-beamforming, their performance remains constrained by the inherent limitations of the delay and sum operation. In Section~\ref{sec:num_exp}, we empirically demonstrate how beamforming accuracy for delay and sum deteriorates when using fractional delay filters with shorter filter lengths. {It is also worth noting that both the recursive formulation developed in \cite{pruned} and the factorization in \cite{vander1} hinge on uniform sensor placement. As we will discuss in Section~\ref{sec:nufft_1}, our algorithm doesn't suffer from this limitation and can easily be adapted to non-uniformly placed sensor scenarios. }
 


\section{Formulation}
\label{sec:form}
We use the standard model for a modulated bandlimited signal impinging on an array of sensors \cite{delude2023slepianbeamformingbroadbandbeamforming,delude2025br}.  An $\Omega$-bandlimited signal $s(t)$ is modulated to a carrier frequency $f_c$ to produce $s_{\mathrm{mod}}(t) = e^{j2\pi f_c t}s(t)$ so that $s_{\mathrm{mod}}(t)$ is spectrally supported over the frequency range $[f_c - \Omega, f_c + \Omega]$.  The modulated signal arrives at the array as a plane wave from angle $\theta = (\varphi, \phi)$, where $\varphi$ is the azimuth and $\phi$ is the elevation relative to the array phase center.  The $m^{\text{th}}$ array element observes a delayed and noisy version of $s_{mod}(t)$; after demodulating, it observes 
\begin{align*}
    y_m(t) &= e^{-j2\pi f_c t}y_{\mathrm{mod},m}(t)\\
    &= e^{-j2\pi f_c \tau_{\theta,m}}s(t - \tau_{\theta,m}) + \text{noise}.
\end{align*}
The $\tau_{\theta,m}$ is the delay for sensor $m$ (relative to the array phase center) and is given by
\[
	\tau_{\theta,m} = \vp_m^T\vartheta(\theta),\ \vartheta(\theta) = \begin{bmatrix}
	    \cos\phi \cos \varphi\\
        \cos\phi \sin \varphi\\
        \sin \phi
	\end{bmatrix}
\]
where $\vp_m$ is the position of the sensor in the standard coordinate system and $\vartheta(\theta)$ is the normal vector in the direction of the incoming plane wave signal.  The demodulated signal $y_m(t)$ is sampled uniformly every $T_{s} \leq \frac{1}{2\Omega}$ seconds at locations $t_n = nT_s$ to produce the sequence
\begin{align}
	\label{eq_1}
    	y_{m}[n] &= e^{-j2\pi f_c \tau_{\theta,m}}s(t_n - \tau_{\theta,m}) + \text{noise},
\end{align}
where $n$ is the sample index.  We refer to the ensemble of $M$ sensor outputs for fixed sample index $n$ as a ``snapshot" of the array. 

We form a beam $b_\theta(t)$ by coherently combining all of the sensor outputs.  The beam $b_\theta(t)$ can be thought of as an estimate of $s(t)$ under the hypothesis that it is arriving from angle $\theta$.  If $s(t)$ is indeed arriving from $\theta$ and there are no other signals present and the noise is additive white Gaussian, then the optimal estimate can be formed using ``delay and sum'':
\[
	b_\theta(t) = \frac{1}{M}\sum_{m=1}^M e^{j2\pi f_c \tau_{\theta,m}}y_m(t+\tau_{\theta,m}).
\]
The key computational step in the above is computing samples of the delayed signals $y_m(t+\tau_{\theta,m})$ from the array samples $y_{m}[n]$.  This is typically done by applying a $K$-tap digital filter with weights\footnote{We have assumed for simplicity that $K$ is even and have absorbed the  $e^{j2\pi f_c \tau_m(\theta)}/M$ into the weights.} $c_{\theta,m}[k]$ that depend on $\tau_{\theta,m}$:
\begin{equation}
	\label{eq:dasapprox}
	b_\theta(t_n) \approx \sum_{m=1}^M\sum_{k=1}^{K} c_{\theta,m}[k] y_{m}[n-k+K/2].
\end{equation}
There are two drawbacks to this approach.  First, if we want to compute beams for $B$ different values of $\theta$, then the total computational cost is $\mathcal{O}(KBM)$ per sample.  For large arrays where $B$ and $M$ are in the 1000s and bandwidths of multiple GHz, this computational cost is significant.  Second, when the $y_m(t)$ are close to critically sampled ($T_s\approx 1/2\Omega$), the finite-length filters introduce a significant bias, even for moderately large $K$ (see Figures \ref{fig:ula_performance} and \ref{fig:upa_performance} below).

Other interpolation schemes have smaller bias than delay and sum but similar computational complexity.  As we see from \eqref{eq_1}, the array samples across multiple snapshots are non-uniform samples of $s(t)$.  We could form a sample of $b_\theta(t_n)$ by taking the $M$ samples closest to $t_n$ (the $M$ indices $(n',m)$ where $|t_n - t_{n'}+\tau_{\theta,m}|$ is smallest) and applying a carefully chosen set of weights to the corresponding samples.  As the non-uniform samples are very dense, there would be almost no bias for this type of estimator.  The computational complexity, however, remains at $\mathcal{O}(BM)$ per sample for forming $B$ beams.

Our approach to beamforming is to estimate multiple samples of $b_\theta(t_n)$ at the same time by solving a linear inverse problem using least-squares.  This approach was formulated and analyzed in detail in \cite{delude2023slepianbeamformingbroadbandbeamforming}, where it was shown to remove almost all the bias inherent to delay and sum while having comparable computational complexity.  In this paper, we show that changing the way the signal is represented allows us to compute many beams simultaneously at an aggregate cost (within logarithmic factors) of $\mathcal{O}(B+M)$, far smaller than the $\mathcal{O}(BM)$ required for delay and sum, while again completely eliminating the bias caused by the finite-length filters.


\begin{figure*}[htbp]
    \centering
    \subfigure[]{
        \includegraphics[width=0.35\textwidth]{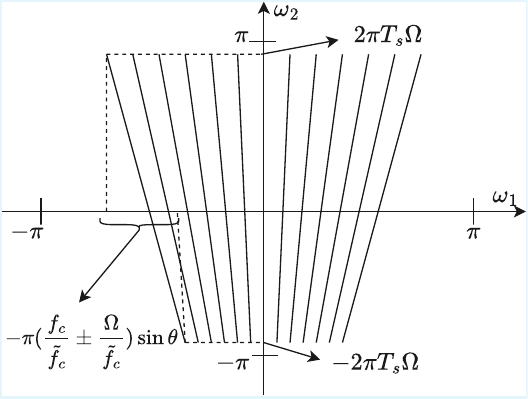}
        \label{fig:single_beam}
    }
    \subfigure[]{
        \includegraphics[width=0.35\textwidth]{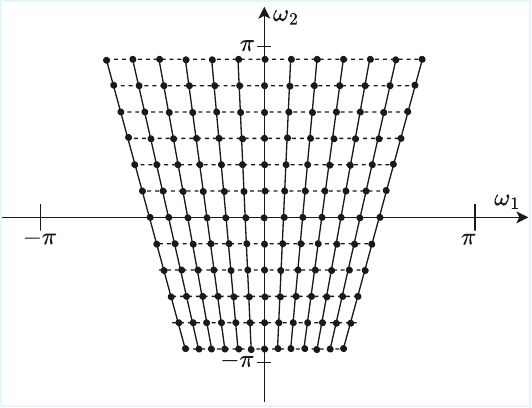}
        \label{fig:multiple_beam}
    }
    \caption{\sl The axes above are in the 2D discrete-time Fourier transform domain; the horizontal $\omega_1$ axis is the
 spatial frequency (Fourier transform across array elements $m$), the vertical $\omega_2$ axis is the frequency in time (Fourier
 transform across samples $n$). (a) The array measurements $\{y_{m,n}\}$ for a signal bandlimited to $\Omega$ at carrier
 frequency $f_c$ are in the linear span of discrete-time sinusoids with frequencies along the slanted line above. The slope of each slanted line depends on the signal bandwidth ($\Omega$) relative to $\tilde{f}_c = f_c + \Omega$ and the direction of arrival; therefore, with $B$ beams, there are $B$ distinct slanted lines in the 2D frequency domain. For representation here $T_s < \frac{1}{2\Omega}$, so we take $\epsilon = 1$. In the general scenario where $T_s = \frac{1}{2\Omega}$, $\epsilon$ is chosen to be just above $1$. This introduces minor aliasing, but it is negligible and outweighed by the reduction in bias gained from accurately capturing the corner frequencies. (b) A pseudo-polar sampling grid, the points are equispaced along the $\omega_2$ axis, and each of the slanted lines is equispaced in slope.}
\end{figure*}

\subsection{Forming a single beam}
We start by describing how our formulation can be used to form a single beam.  We then discuss how the key computational step in this approach can be executed across many beams simultaneously.

\subsubsection{Representation by Fourier extension}
Our algorithm will work from $N$ array snapshots at times $\{t_n\}_{n=1}^N$ and produce $N$ uniform samples (with the same spacing $T_s$) for $B$ beams.  For a single beam at angle $\theta$,  we treat the $MN$ array samples as phase-modulated non-uniform samples  of a signal $s(t)$ arriving from $\theta$.  These samples are at known delays $\{t_n-\tau_{\theta,m}\}_{m,n}$ and lie in a time interval of length
\[
	T_\theta = \max_{m}\{\tau_{\theta,m}\} - \min_{m}\{\tau_{\theta,m}\} + (N-1)T_s.
\]
We will represent the signal over this time interval using a linear combination of sinusoids\footnote{We assume $N$ and $L$ to be even, again for convenience of expoistion.} from an overcomplete basis:
\begin{equation}
	\label{eq:stfourierapprox}
	s(t) \approx \sum_{\ell=1}^{L} \beta_\ell \e^{j \omega_\ell t},
	\quad \omega_\ell =2\pi \epsilon\frac{ \ell - \frac{L}{2}}{L T_s},
	\quad L = \gamma N ,
\end{equation}
where $\gamma > 1$ and $\epsilon > 1$ are small expansion factors.  This represents the signal using $L > N$ sinusoids in the frequency range
\[
    \epsilon\left[-\frac{1}{2T_s},\frac{1}{2Ts}\right]
    \subseteq \epsilon[-\Omega,\Omega].
\]
%
%
For critically sampled signals with $T_s = \tfrac{1}{2\Omega}$, taking $\epsilon > 1$ ensures that the ``corner frequencies'' are represented with the same accurracy as the frequencies on the interior, thereby avoiding some bias in the estimation process. When $T_s < \tfrac{1}{2\Omega}$, we can simply take $\epsilon = 1$. 

As we will use the $\{\beta_\ell\}$ to reconstruct $N$ samples of the signal, the parameter $\gamma$ controls the 
overcompleteness the representation. 
If $\gamma = \epsilon \frac{T_\theta}{N T_s}$, then \eqref{eq:stfourierapprox} is an orthogonal Fourier series expansion over a time interval of lenght $T_\theta$.  For 
\begin{equation}\label{eq:gamma_lb}
    \gamma > \epsilon \frac{T_\theta}{N T_s},
\end{equation} 
the $\{\e^{j\omega_\ell t}\}_\ell$ are an overcomplete Fourier basis, often referred to as a \emph{Fourier extension} series with extension factor of 
\[
    \gamma' = \frac{\gamma}{\epsilon} \frac{N T_s}{T_\theta}.
\]
In this case, there are more basis functions ($> 2\epsilon\Omega T_\theta$) than there are degrees of freedom in a bandlimited signal on an interval of length $T_\theta$ ($\approx 2\Omega T_\theta$). 
This overcompleteness is critical for the approximation accuracy, as the standard Fourier series has well-known shortcomings (Gibbs phenomena, ringing) when used for bandlimited signals over a finite time window.
Recent work \cite{adcock2012resolutionpowerfourierextensions, adcock2013numericalstabilityfourierextensions, DAAN} has shown that for modest extension factors ($\gamma' = 2$, say), there exist $\{\beta_\ell\}$ so that the approximation error (for a bandlimited signal) decays exponentially as terms are added.  


The overcompleteness also means that the $\{\beta_\ell\}$ used in \eqref{eq:stfourierapprox} are in general non-unique: there will be many $\{\beta_\ell\}$ that give essentially the same approximation.  This issue is easily addressed using standard regularization, as we will see below.

We note that while there are multiple ways to efficiently represent a bandlimited signal over a finite length of time (for example, \cite{delude2023slepianbeamformingbroadbandbeamforming, delude2025br} use orthogonal Slepian functions), using overcomplete sinusoids will allow us to accelerate the computations.  

\vspace{.1in}
\subsubsection{Forming the beam using least-squares}
With this representation in hand, we can now set up the inverse problem to solve to form a beam.  The collection of $N$ snapshots gives us a total of $MN$ samples that we collect into the vector $\vy$.  For each beam angle $\theta$, we estimate the corresponding expansion coefficients $\vbeta_\theta$ in \eqref{eq:stfourierapprox} using the regularized least-squares estimate
\begin{equation}
	\label{eq:betahatls}
	\hat\vbeta_\theta  = \left(\mF_\theta^\H\mF_\theta+\delta\mId\right)^{-1}\mF_\theta^\H\vy,	
\end{equation}
where $\mF_\theta$ is a $MN\times L$ non-uniform Fourier matrix whose row index $m'=1,\ldots,MN$ can be associated with a (sensor, snapshot) pairing $(m,n),m=1,\ldots,M,~n=1,\ldots,N$ so that 
\begin{equation}
	\label{eq:FthetaH}
	\left[\mF_\theta\right]_{m',\ell} = \e^{-j2\pi f_c\tau_{\theta,m}}\cdot\e^{j\omega_\ell(t_n-\tau_{\theta,m})}.
\end{equation}
The $N$ beam samples can then be recovered by applying the inverse Fourier reconstruction matrix $\mF_u$
\begin{equation}
	\label{eq:bhatls}
	\hat\vb_\theta = \mF_u\hat\vbeta_\theta,\quad \left[\mF_u\right]_{n,\ell} = \e^{j\omega_\ell t_n}.
\end{equation}
The matrix $\mF_\theta^\H\mF_\theta+\delta\mId$ to invert in \eqref{eq:betahatls} is small (as $L =\gamma N\ll MN$) and has Toeplitz structure.  For small values of $N$, it can be pre-computed and applied at cost $\mathcal{O}(N^2)$.  For large values of $N$, it is possible to apply the inverse in $\mathcal{O}(N\log N)$ time using a series of FFTs; this is discussed more in Section~\ref{sec:fi} below.  There is also a principled way to choose the regularization parameter $\delta$ which we will discuss for the numerical experiments in Section~\ref{sec:num_exp}.  

\vspace{.1in}
\subsubsection{Choosing the number of snapshots $N$}
The number $N$ of snapshots we want to process simultaneously depends on the sampling interval $T_s$ (and hence on the bandwidth $\Omega$) and the dimensions of the array aperture.  Choosing $N$ too small will prevent us from taking advantage of the coherence between snapshots present in the broadband setting.  As $N$ increases, the and memory needed in processing the snapshots grows linearly while the computation needed per sample grows logarithmically.

To estimate the signal at any particular time $t_0$, we want to use all of the snapshots that contain samples around $t_0$.  As when estimating a signal $s(t)$ from direction $\theta$ each snapshot carries samples in an interval of length $\max_m\{\tau_{\theta,m}\}-\min_m\{\tau_{\theta,m}\}$, we will need
\begin{equation}
    \label{eq:N_lb}
    N > \max_{\theta}\left(\frac{\max_m\{\tau_{\theta,m}\}-\min_m\{\tau_{\theta,m}\}}{T_s}\right).
\end{equation}
In general, taking $N$ to be roughly twice the right-hand side of the above saturates the accuracy of the approximation.  This is the guideline we use in the numerical experiments in Section~\ref{sec:num_exp}.

For standard array geometries, the lower bound in \eqref{eq:N_lb} will depend on the number of array elements $M$.  For example, if we fix $f_c$ and $\Omega$ and use half-wavelength spacing for the array elements, the effective aperture $\max_m\{\tau_{\theta,m}\}-\min_m\{\tau_{\theta,m}\}$ will grow proportional to $M$ for a uniform linear array and proportional to $\sqrt{M}$ for a uniform planar array.  These affects are carefully accounted for when we discuss how our computations scale, but in many practical scenarios, $N$ can be thought of as a small constant --- in our numerical experiments, there is not a significant difference in runtime between our ``FBST Precompute'' methods whose per-sample complexity scales $\mathcal{O}(BN)$ and our ``FBST Superfast'' methods that reduce this to $\mathcal{O}(B\log N)$.


Finally, while our formulation and discussion throughout the paper is centered on estimating a single batch of data, the framework can also be used to process array measurement in a streaming manner.  It is straightforward to adapt the analysis and algorithms in the following sections to handle contiguous and possibly overlapping batches of snapshots that are acquired sequentially in time.

\subsection{Forming multiple beams}\label{subsec:fmb}

The beamformer given by \eqref{eq:betahatls} and \eqref{eq:bhatls} solves for beam samples at a single angle $\theta$.  Ultimately, this requires the application of an $N\times MN$ matrix (that can be precomputed for a set of prescribed angles) to the $MN$-vector $\vy$.  Though there are many ways that the computations can be accelerated using non-uniform FFTs and fast Toeplitz solvers, constructing each of the $B$ beams separately will require at least $\mathcal{O}(BM)$ operations per sample.  Implemented naively, the beamformer is more accurate than the approximate delay and sum in \eqref{eq:dasapprox} but has the same order of computational complexity.

In the next two sections, we will show that $B\sim M$ beams can be computed \emph{simultaneously} using significantly fewer computations per sample.  We can choose a set of angles $\{\theta_b\}_{b=1}^B$ that give us full coverage over the aperture and allow us to compute $\mF_{\theta_b}^\H\vy$ for all $b=1,\ldots,B$ using a single fast-Fourier-like operator in $\mathcal{O}(MN\log N + BN\log B)$ time for a per-sample cost of $\mathcal{O}(M\log N + B\log B)$.  

Following this, each $\hat\vbeta_{\theta_b}$ cab be solved for independently and in parallel.  As noted above, each of these systems is relatively small, $\mathcal{O}(N\times N)$, and has Toeplitz structure.  There are two alternatives: the $N\times L$ matrices $\mF_u(\mF_{\theta_b}^\H\mF_{\theta_b})^{-1}$ can be pre-computed for all $\theta_b$ and then simply applied to the $\mF_{\theta_b}^\H\vy$ at a per-sample computational cost of $\mathcal{O}(BN)$ and a storage cost of $\mathcal{O}(BN^2)$ for all $B$ beams.  Alternatively, the inverses $(\mF_{\theta_b}^\H\mF_{\theta_b})^{-1}$ can be decomposed as four linear convolutions with pre-computed vectors (this is detailed in Section~\ref{sec:fi}) for a per-sample computational cost of $\mathcal{O}(B\log N)$ and a storage cost of $\mathcal{O}(BN)$.  
In theory, we will need $N$ to be a small fraction of $M$ for a linear array and as a small fraction of $\sqrt{M}$ for a planar array according to \eqref{eq:N_lb}, making the difference between the scalings $\mathcal{O}(BN)$ and $\mathcal{O}(B\log N)$ crucial to our asymptotics.  For practical array geometries the value of $N$ is moderate, and the direct application of $\mF_u(\mF_{\theta_b}^\H\mF_{\theta_b})^{-1}$ may be faster than the multiple convolutions (this is indeed the case in our numerical experiments below where we take $N=64$).

\section{A Fan FFT for beamspace Projections}
\label{sec:transform}

In this section, we describe how to compute the $\{\mF_{\theta_b}^\H\vy\}_{b=1}^B$ simultaneously in $\mathcal{O}(MN\log N + BN\log B)$ time using a ``fan shaped'' fast Fourier transform.  We will focus our exposition on the special case of the uniform linear array (ULA), the extension to the uniform planar array (UPA) can be found in Appendix~\ref{appdx:fbst_upa}.

We will assume our ULA has half-wavelength spacing corresponding to the maximum frequency $f_c+\Omega$, so $\lambda/2 = c/2(f_c+\Omega)$.  For a signal arriving $\theta$ degrees off broadside, this makes the delays
\[
	\tau_{\theta,m} = \frac{\sin\theta}{2(f_c+\Omega)}m,\quad m=0,\ldots,M-1.
\]
With these values of delays and using $t_n = nT_s$, we can write the $\ell^{\text{th}}$ entry of $\mF_\theta^\H\vy$ as
\begin{align}\label{eq:sum_1}
	\left[\mF_\theta^\H\vy\right]_\ell &= \sum_{n=1}^N\sum_{m=1}^My_m[n]\e^{j(2\pi f_c\tau_{\theta,m}-\omega_\ell t_n + \omega_\ell\tau_{\theta,m})} \\
	&= \sum_{n=1}^N\sum_{m=1}^My_m[n]\e^{-j\omega_1(\ell)m}\e^{-j\omega_2(\ell)n}
\end{align}
where
\begin{align*}
    \omega_1(\ell) &= -\frac{\pi\sin\theta}{f_c+\Omega}\left(f_c + \frac{\epsilon}{LT_s}\cdot(\ell-\frac{L}{2})\right)\\
	\omega_2(\ell) &= \frac{2\pi\epsilon}{L}\cdot(\ell - \frac{L}{2})
\end{align*}
This means that $\mF_\theta^\H\vy$ is a collection of $L$ equally spaced samples along a diagonal line of the 2D discrete time Fourier transform (DTFT) of $\{y_m[n]\}_{m,n}$, as shown in Figure~\ref{fig:single_beam} --- $\omega_1(\ell)$ is the spatial frequency for variations across the array, while $\omega_2(\ell)$ is the temporal frequency for variations in the signal across time. When forming multiple beams we will have several of these slanting lines as depicted in Figure~\ref{fig:single_beam} with slope depending on $\theta$.
If we choose beam angles such that the corresponding lines are equispaced in slope, then we get a fan-like pattern in the 2D DTFT domain as illustrated in Figure \ref{fig:multiple_beam}. For $B$ odd, we can take
\begin{equation}
    \label{eq:thetab}
    \theta_b = \arcsin \frac{2b -B-1}{B-1}, \ b=\{1,2,\ldots,B\}
\end{equation}
to cover the range $\theta\in[-\pi/2,\pi/2]$; for $B$ even, we can take $\theta_b = \arcsin{\left((2b-B+1)/B\right )}$.


Similar fan-like Fourier sampling patterns arise often in imaging problems \cite{7488260, 7950465}.  In these applications, the Fourier samples can be computed efficiently using the ``pseudo-polar Fourier transform'' \cite{AVERBUCH2006145}. The same underlying tool that makes pseudo-polar Fourier transforms efficient, the fast chirp $z-$ transform \cite{6772159}, can be used to evaluate the Fourier samples in \eqref{eq:sum_1} shown in Figure~\ref{fig:multiple_beam}.


For the $\{\theta_b\}$ in \eqref{eq:thetab} we can rewrite the summation in \eqref{eq:sum_1}
\begin{align}\label{eq:DFT_1}
    w_b(\ell) = \left[\mF_{\theta_b}^\H\vy\right]_\ell &= \sum_{m,n}y_{m,n}e^{j\pi(\zeta + \xi \ell' )b' m}e^{-j\frac{2\pi\epsilon}{L}\ell' n},
\end{align}
where we have $\zeta = \frac{2f_c}{(B-1)(f_c+\Omega)}$, $\xi = \frac{2\epsilon}{(B-1) L T_s (f_c+\Omega)}$, $\ell' = \ell - \frac{L}{2}$ and $b' = b - \frac{B+1}{2}$.  To compute the above, we start by computing
\[
    \hat{y}_{m,\ell} = \sum_{n}y_{m,n}z_{\ell'}^{-n},
    \quad z_{\ell} = \e^{-j2\pi\epsilon\ell'/L},
\]
for each $m=1,\ldots,M$.  This amounts to $M$ chirp $z$'s mapping $N$ time samples to $L$ frequency samples, costing a total of $\mathcal{O}(MN\log N)$ (since $L$ is a small multiple of $N$).  We then compute 
\[
    w_{b,\ell} = \sum_{m}\hat{y}_{m,\ell}z_{b}^{-m},
    \quad z_{b} = e^{-j\pi(\zeta + \xi \ell' )b'}, 
\]
for each $\ell=1,\ldots,L$.  This amounts to $L$ chirp $z$'s mapping $M$ array samples to $B$ frequency samples, costing a total of $\mathcal{O}(NB\log B)$.
%
%
Making the obvious index change, all $BL$ computations of $w_b(\ell)$ have $\mathcal{O}(MN\log{}N + BN\log{}B)$ complexity, or $\mathcal{O}(M\log N + B\log B)$ per sample.

\section{Fast Inverse}
\label{sec:fi}
With the beams $\{\vw_b\}_{b=1}^B$ in hand, the corresponding expansion coefficients $\{\hat{\vbeta}_{b}\}_{b=1}^B$ can be estimated as,
\[
\hat{\vbeta}_{b} = \left(\underbrace{ \mF_b^H\mF_b + \delta\mId }_{\mA_b}\right)^{-1}\vw_b.
\]
As alluded to before $\mA_b$ has a Toeplitz structure. To see this we write each entry as,
\[
\left[\mA_b\right]_{\ell,k} = \sum_{m'}e^{(\omega_{\ell}-\omega_k)t_m'} + \delta\RN{1}(\ell -k )
\]
where $\RN{1}(\cdot)$ is the indicator function and $\{m'\}_{m'=1}^{MN}$ is the mapping introduced before \eqref{eq:FthetaH}. Since the $\{\omega_\ell\}_{\ell=1}^{L}$ (given in \eqref{eq:stfourierapprox}) are spaced uniformly, the entries $\left[\mA_b\right]_{\ell,k}$ are function of $(\ell -k)$, and so the matrix is Toeplitz. Although the inverse of a Toeplitz matrix is not Toeplitz itself, it is persymmetric and can be recovered using its first and last columns using the Gohberg-Semencul formula,
\[
\mA_b^{-1} = \mL_{\vx}\mU_{\vy} - \mL_{\vy}\mU_{\vx}.
\]
In the above equation $\{\vx,\vy\}$ are the first and last column of $\mA_b^{-1}$ and $\mL$, $\mU$ are lower and upper triangular Toeplitz matrices constructed using these vectors. Therefore, we can express the inverse of $\mA_b$ as a sum of products of Toeplitz matrices. For more details we direct the readers to \cite{HEINIG2002199,HeinigRost+1984,doi:10.1137/S0895479899362302}.

As Toeplitz matrices can be embedded in circulant matrices, the matrix-vector product $\mA_b^{-1}\vw_b$ can be computed using four FFT pairs in $\mathcal{O}(N\log N)$ time, so the $\hat\vbeta_b$ for all $B$ beams can be computed in $\mathcal{O}(BN\log N)$ time. Since we only need to store two generating vectors $\{\vx,\vy\}$ for constructing each inverse, the total space complexity associated with this step is $\mathcal{O}(BN)$.  For a fixed set of beam angles $\{\theta_b\}$, these generating vectors can be precomputed, but we also note that efficient $\mathcal{O}(N\log^2 N)$ algorithms exist for computing these on-the-fly { \cite{doi:10.1137/S0895479899362302}}.

Once the expansion coefficients are recovered, the $N$ uniform samples can be reconstructed in $\mathcal{O}(BN\log N)$ operations by rewriting $\mF_u \vbeta_b$ as a chirp $z-$transform in index $n$.

As highlighted in section \ref{subsec:fmb}, this Toeplitz inversion scheme is especially useful when $N$ is large; for modest values of $N$, it would be more convenient to just evaluate $\mF_u\mA_b$ beforehand for all beams and apply it at runtime in $\mathcal{O}(BN^2)$ operations. Table \ref{table:comparison} tabulates the per-sample runtime complexity of the two variants of FBST and the conventional delay and sum beamformer. Note that for $ B \approx M$ the complexity grows quadratically in $M$ for the delay and sum, whereas it is linear for FBST.

 \begin{table}[]
 \centering
 \caption{Per-sample Runtime complexity for FBST, Delay and Sum.}
\begin{tabular}{|l|l|}
\hline
\textbf{Algorithm} & \textbf{Time complexity}
\\ \hline\hline
FBST (Precompute Inverse) & $\mathcal{O}(M\log N + B\log B + BN)$
\\ \hline
FBST (Superfast Toeplitz) &  $\mathcal{O}(M\log N + B\log B + B\log N)$
\\ \hline
Delay and Sum & $\mathcal{O}(MB)$\\
\hline
\end{tabular}
\label{table:comparison}
\end{table}

\section{Numerical Experiments}
\label{sec:num_exp}

This section presents numerical experiments to test the performance of the proposed algorithm on uniform linear array (ULA) and uniform planar array (UPA) configurations. We use two delay and sum beamformer variants to 
benchmark the beamforming performance and the computational efficiency of our algorithm. The first variant is a conventional delay and sum (DS) implemented using fractional delay filters truncated to $R$ taps. The second is a fast delay and sum (FDS) beamforming technique proposed in \cite{pruned, quadtree}.  Additionally, we compare our algorithm against sub-band processing to evaluate its efficacy against traditional frequency-based beamforming methods.

FDS is a multistage algorithm with a filter bank for interpolation in each stage. We keep the truncated filter lengths in FDS the same across all stages. Similar to the original work on FDS, we employ a radix-2 beamformer with $S = \log_2(M)$ stages. At each stage, FDS downsamples by a factor of 2 in the spatial domain while doubling the number of partial beams formed. Hence, in the final stage, we end up with $2^S = M$ total beams. Both the DS and FDS beamformers use a sinc interpolator as the filtering mechanism. Also note that the FDS presented in \cite{pruned} is specifically designed for a ULA; we extend it to a UPA geometry using the procedure described in the Appendix.

Sub-band processing has the same procedure as alluded to in Section \ref{sec:related}. We first take an FFT across $N$ snapshots. Once in the frequency domain, we use chirp $z$ to speed up beamspace computations across frequency bins. Finally, we take an inverse FFT across all the beams to get the corresponding beams in the time domain.

The incident broadband signal is generated using a sum of sinusoids model\footnote{In this model we form a signal by randomly weighting sinusoids generated at random frequencies within the band.} and corrupted by Gaussian noise. The experiments are configured with a carrier frequency of $f_c = 20$ GHz, bandwidth of $\Omega = 5$ GHz, and a sampling frequency of $f_s = 2\Omega$. For assessing the beamforming performance, we use a $128$ element ULA and a $16 \times 16$ element UPA, with $N=64$ snapshots for both cases. The $\epsilon$ parameter is set to 1.01 for both. {Note that this introduces slight aliasing, but its impact on overall performance is negligible, as it is outweighed by the substantial reduction in bias achieved by properly capturing the corner frequencies within the bandwidth.}
%
%
The regularization parameter $\delta$ in \eqref{eq:betahatls} can be set based on the noise variance and the spectrum of $\mF_\theta$ which is known a priori and is a function of the array geometry.  In practice, the reconstruction process is not overly sensitive to the choice of $\delta$.  In our experiments, we simply fixed $\delta = 10^{-5}$ across a $-30$ dB to $30$ dB SNR test range and results showed near-ideal (e.g., perfectly matching true time delay) performance for a variety of array sizes and geometries.

\begin{figure*}[t]
    \centering
    \subfigure[]{\includegraphics[width=0.23\textwidth]{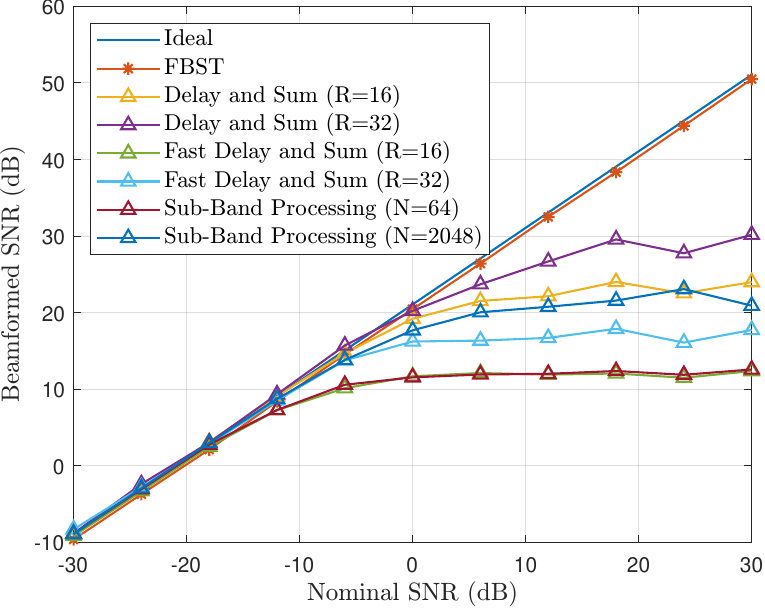} \label{fig:ula_performance}}
    \hfill
    \subfigure[]{\includegraphics[width=0.23\textwidth]{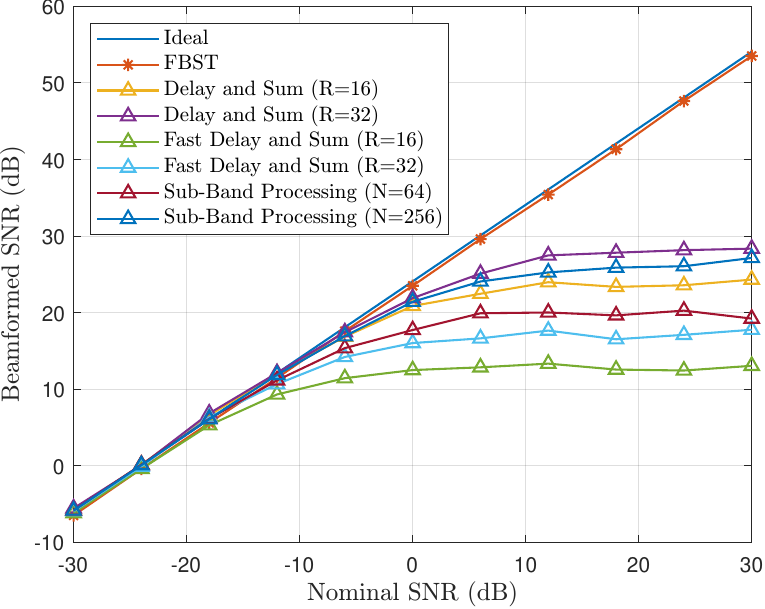} \label{fig:upa_performance}}
    \hfill
    \subfigure[]{\includegraphics[width=0.23\textwidth]{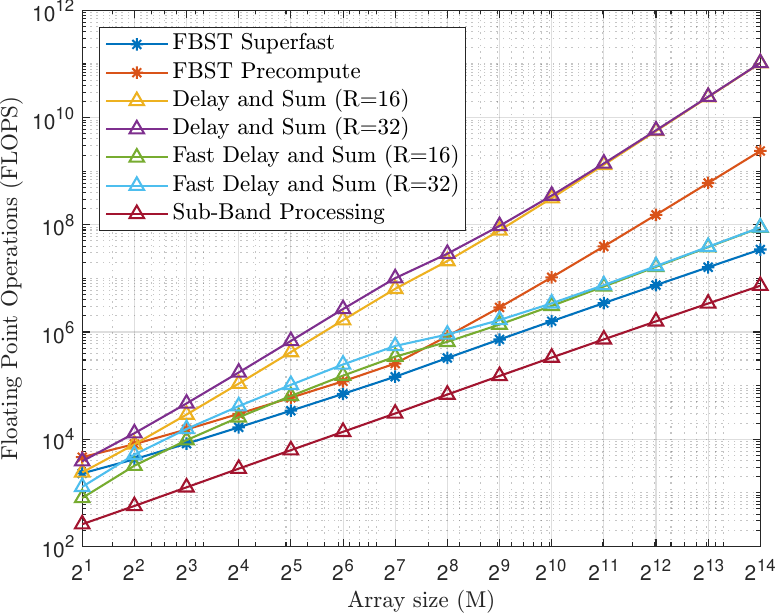} \label{fig:ula_runtime}}
    \hfill
    \subfigure[]{\includegraphics[width=0.23\textwidth]{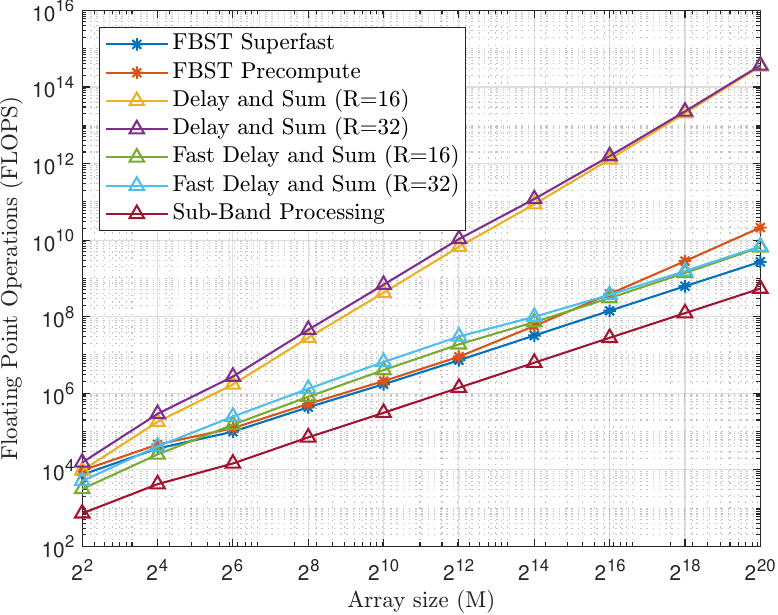} \label{fig:upa_runtime}}
    \caption{\sl (a) Beamformed SNR vs. Nominal SNR for a 128-element ULA with 64 snapshots. (b) Beamformer SNR vs. Nominal SNR for a 16$\times$16-element UPA with 64 snapshots. (c) No. of floating point operations vs. array size for a ULA. The number of beams scales as $B\approx M$ for all algorithms. Two filter lengths $R=16,32$ are used for measuring the DS, FDS computations. For sub-band processing, the number of snapshots used for FFT across time is the same as used in the FBST algorithm and set according to \eqref{eq:N_lb}. (d) No. of floating point operations as a function of array size for a UPA, where the sensors are arranged in a square grid of dimensions \( \sqrt{M} \times \sqrt{M} \). For all algorithms, the number of beams scales approximately linearly with the number of sensors, i.e., \( B \approx M \). Two filter lengths $R=16,32$, are used for measuring the DS, FDS computations. For sub-band processing, the number of snapshots used for FFT across time is the same as used in the FBST algorithm and set according to \eqref{eq:N_lb}.}
    \label{fig:four_subfigure}
\end{figure*}

Based on the given parameters, the lower bounds for $\gamma$ from \eqref{eq:gamma_lb} evaluate to 1.63 for the ULA and 1.11 for the UPA. Therefore, we set $\gamma = 2$ for both experiments. An on-grid angle is selected to assess the beamforming performance for both array types. Figure \ref{fig:ula_performance} illustrates the beamformed SNR versus nominal SNR for the ULA at $\theta \approx 61^{\circ}$, while Figure \ref{fig:upa_performance} depicts the same for the UPA at $(\varphi,\phi)\approx (57^{\circ},77^{\circ})$. The results demonstrate that the FBST algorithm effectively attains the ideal array gain, even at higher SNRs, as it closely tracks the ideal beamformed SNR (depicted in blue). 

In contrast, delay and sum beamformers experience saturation at higher SNRs due to the truncation of the interpolating filter. This induces a bias that begins to dominate the error as the noise variance decreases. This truncation effect is even more pronounced in the FDS beamformer because each filtering stage induces its own bias. Hence the bias accumulates across stages, and results in a significantly higher final-stage bias\footnote{This phenomena is readily observed in the original computational experiments provided in \cite{pruned,quadtree}.}. Consequently, the beamforming performance of FDS is poorer than that of conventional delay and sum beamforming and will continue to degrade with additional stages, e.g., with larger arrays. 

A similar conclusion can be drawn regarding the beamforming performance of sub-band processing. When only a small number of FFT snapshots are used, the resulting coarse frequency resolution leads to degraded beamforming performance. This is because, as with any finite $N$, the phase adjustment at each frequency only induces an approximate delay.
As seen in Figures~\ref{fig:ula_performance} and~\ref{fig:upa_performance}, increasing the number of snapshots can help alleviate this issue, but it comes at the expense of a larger buffer size and increased latency --- even with buffer sizes in the 1000s, the limited frequency resolution introduces a bias that dominates the performance at medium SNRs.

Figures \ref{fig:ula_runtime} and \ref{fig:upa_runtime} illustrate the total floating point operations (FLOPS) per-sample of various algorithms as the array size increases. The total number of beams also scales with array size, while $\gamma=2$ is kept constant and $N$ is set according to \eqref{eq:N_lb}. This implies that for ULA $N$ scales as $\mathcal{O}(M)$ and for UPA it scales as $\mathcal{O}(\sqrt{M})$. Thus $N$ is typically much smaller for a UPA, which leads to FBST precompute being almost as fast as FBST superfast. The results indicate that FBST-based methods exhibit better scalability compared to conventional DS beamforming. Although the operations in FDS and sub-band processing are similar in scale to FBST Superfast, this is negated by their relatively poor beamforming performance.

The experiments presented above are based on the sequential execution of all algorithms. However, it is important to emphasize the parallelization potential of the FBST algorithm. Specifically, during the transform step, the summations in \eqref{eq:DFT_1} can be independently computed for each pair of indices $l,b$, enabling parallel execution. Likewise, as previously mentioned, the inverse problem step can also be parallelized across different beams. In contrast, such parallelization is not feasible for the FDS algorithm due to its inherently recursive structure, which necessitates a sequential computation.

Finally, Figure \ref{fig:ula_beampattern} plots the beam pattern for FBST, DS, FDS and sub-band processing for the ULA parameters mentioned earlier. To capture the response across the entire frequency band, beam patterns are computed at 20 evenly spaced frequencies within $[f_c-\Omega,f_c+\Omega]$ and then overlaid to illustrate how the beamformer responds across the entire range. We can observe from the inlaid plot that the large bias in FDS beamforming results in frequency-dependent fluctuations in the steered direction. In sub-band processing, insufficient frequency sampling causes beam squint, i.e. different frequency components are steered toward slightly different directions. In contrast, the low bias in FBST yields a stable response in the intended steering direction across all frequencies, maintaining consistency and avoiding beam squint.

In the following section, we present an analysis of the reconstruction error when using the Fourier extension basis to recover the least squares estimate for a band-limited signal.

\section{Error Analysis for Fourier extension}
\label{sec:err}

One of the benefits of recasting broadband beamforming as a linear inverse problem is that we can use well-established tools from linear algebra to analyze and predict its performance in an extremely precise manner.
In this section, we analyze the reconstruction performance of the proposed broadband beamformer by decomposing the estimation error into its \textit{bias} and \textit{variance} components. Our goal is to understand how well the Fourier extension basis can approximate signals that are naturally represented using the Slepian functions \cite{delude2023slepianbeamformingbroadbandbeamforming} and to quantify how this approximation improves with the design parameters---particularly the extension factor~$\gamma$, and the number of array snapshots~$N$. We show that a (very small) bias arises from model mismatch between the (finite) Fourier and (infinite) Slepian representations. At the same time, the variance reflects the effect of the measurement noise propagated through the regularized least-squares estimator. The key results demonstrate that: 1) the bias rapidly decreases once $\gamma$ exceeds the theoretical lower bound in~\eqref{eq:gamma_lb}, and 2) the variance approaches the ideal array gain of~$\sigma^2/M$ as~$N \to \infty$. Together, these findings establish the Fourier extension approach as an accurate and computationally efficient approximation to the Slepian representation for broadband beamforming.

At a high level, each beam is formed by estimating an $\Omega$-bandlimited signal on an interval $[0,T]$ from samples at locations $\{t_{m'}\}_{m'=1}^{MN}$. The samples are non-uniform but very dense.

The signal can be represented exactly using an infinite number of Slepian coefficients $\{\alpha_n\}_{n=1}^{\infty}$,
\[
s(t) = \sum_{n=1}^{\infty}\alpha_n\phi_n(t), t\in [0,T],
\]
where the $\{\phi_n\}$ are the eigenfunctions of the integral operator with kernel
\begin{equation}\label{eq:operator}
k(t,s) = \frac{\sin{\Omega(t - s)}}{\pi(t - s)} = \sum_{n=1}^{\infty}\lambda_n\phi_n(t)\phi_n(s).
\end{equation}
A detailed overview of this Slepian decompostion and its application in broadband beamforming can be found in \cite{delude2023slepianbeamformingbroadbandbeamforming,delude2025br ,book_slepian, MOORE2004208}. We will also find it helpful to write $s = \mathcal{S}\valpha$, where $\mathcal{S}:l_2 \rightarrow L_2([0,T])$ maps the Slepian coefficients into the continuous-time signal. If we conceptualize $\mathcal{S}$ as a matrix that has an infinite number of columns and each column is a continuous-time function, then the entries of $\mathcal{S}$ are $\mathcal{S}_{t,n} = \phi_n(t)$.

Our observational model for the samples generated at the array is 
\[
\vy = \mathcal{A}\valpha + \vn
\]
where $\mathcal{A}:l_2\rightarrow \mathbb{R}^{MN}$ is the linear operator that takes the Slepian coefficients and returns the samples at $\{t_{m'}\}$ and $\vn \sim \mathcal{N}(0,\sigma^2\mId)$ is the zero mean Gaussian noise.  If we, again, conceptualize this as a matrix with $MN$ rows and an infinite number of columns,
then the entries are $\mathcal{A}_{m',n} = \phi_n(t_{m'})$. 

Our fast beamformer approximates the signal by estimating the coefficients in an extended Fourier dictionary with sinusoids at frequencies $\{\omega_{\ell}\}_{\ell=1}^L$ given in \eqref{eq:stfourierapprox}. It computes $\hat{\vbeta}$ according to \eqref{eq:betahatls} and takes
\[
\hat{\vs} = \mathcal{F}^*\hat{\vbeta}.
\]
where the linear operator $\mathcal{F}^* : \mathbb{C}^{L} \rightarrow L_2([0,T])$ has entries $\mathcal{F}^*_{t,\ell} = e^{j \omega_{\ell} t}$.

\begin{figure*}[t]
    \subfigure[]{
    \includegraphics[width=0.23\linewidth]{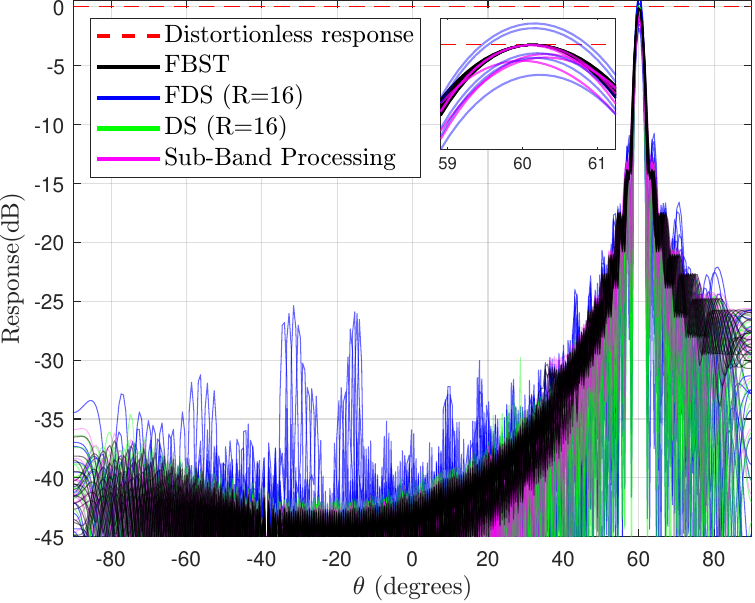}
    \label{fig:ula_beampattern}
    }
    \subfigure[]{
    \includegraphics[width=0.23\linewidth]{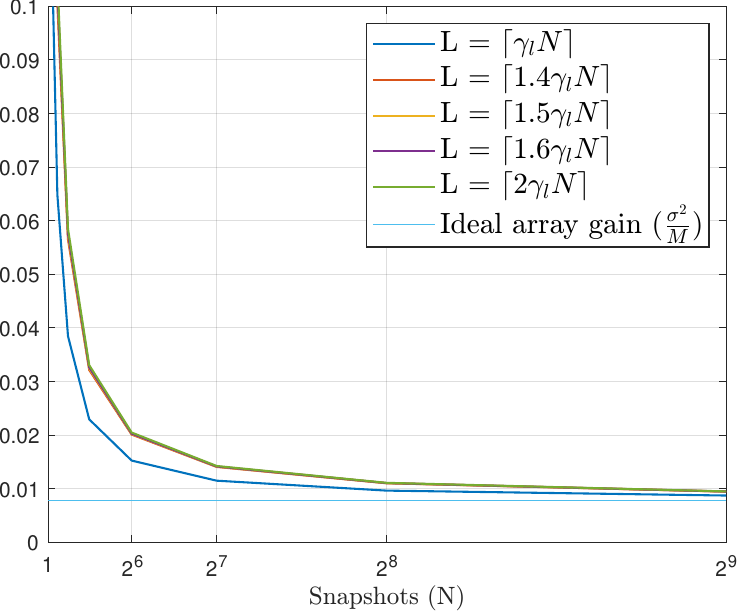}
    \label{fig:variance_vs_snr}
    }
    \subfigure[]{
    \includegraphics[width=0.23\linewidth]{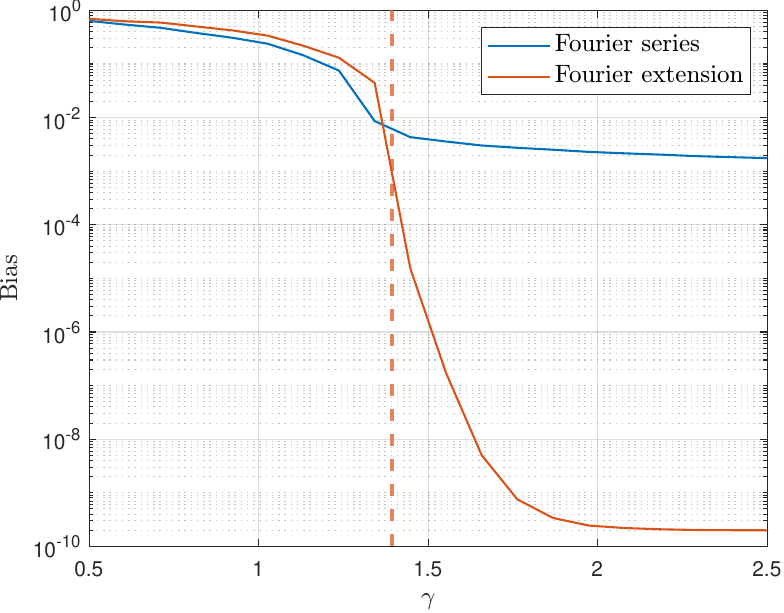}
    \label{fig:bias_vs_gamma}
    }
    \subfigure[]{
    \includegraphics[width=0.23\linewidth]{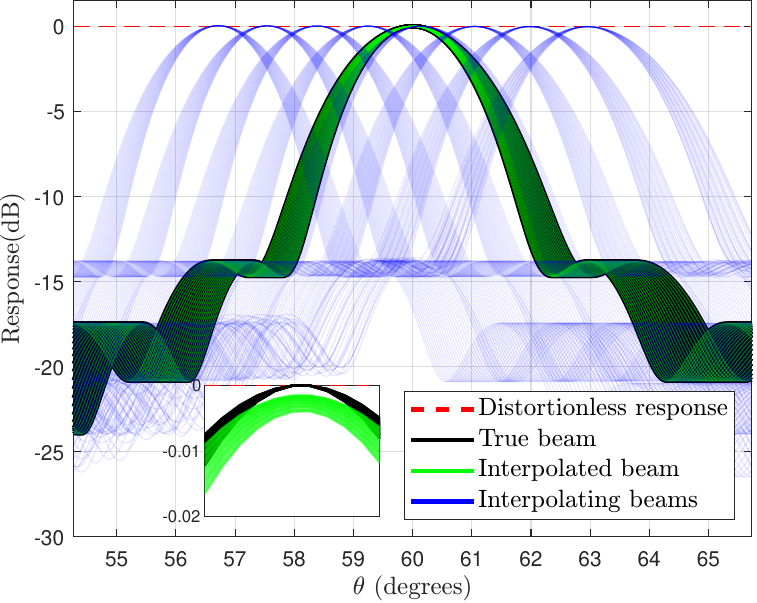}
    \label{fig:super_res}
    }
    \caption{\sl (a) Beam patterns for the FBST, DS, FDS and sub-band processing algorithms on a 128-element ULA with 64 snapshots. $R=16$ tap length is used for DS and FDS. Beam patterns for 20 frequencies in $[f_c - \Omega, f_c + \Omega]$ are superimposed to get frequency response over the entire band.
    The inlaid plot provides a zoomed-in view of the FBST, FDS and sub-band processing responses around the steered direction for a subset of frequencies to highlight beam squint in sub-band processing and gain-fluctuations in FDS. (b) Variance vs. number of snapshots $N$ plotted for five values of $L$. As $N \rightarrow \infty$, the variance approaches the ideal array gain. (c) Bias vs. $\gamma$ plotted for Fourier extension and Fourier series. Fourier extension has negligible bias once we go over the bound given in \eqref{eq:gamma_lb} by a modest amount. (d) Zoomed-in beam pattern for interpolated and true beams on a 128-element ULA. A total of 256 beams were formed, out of which only 8 were used for interpolating to an off-grid angle $\theta = \frac{\pi}{3}$. Beam patterns for 20 frequencies in $[f_c - \Omega, f_c + \Omega]$ are superimposed to get the frequency response over the entire band.}
\end{figure*}

The reconstruction error for this process can be broken into two parts 
\begin{align*}
    \hat{\vs} - \vs &= \underbrace{(\mathcal{F}^*(\mF^H\mF + \delta\mId)^{-1}\mF\mathcal{A} - \mathcal{S})\valpha}_{\text{bias}}\\
    &+\underbrace{\mathcal{F}^*(\mF^H\mF + \delta\mId)^{-1}\mF\vn}_{\text{variance}}.
\end{align*}
For a given set of sample locations $\{t_{m'}\}_{m'=1}^{MN}$, we can compute both of these as follows.
The variance term is a Gaussian random process with covariance kernel
\[
\vr = \sigma^2 \mathcal{F}^*\underbrace{(\mF^H\mF + \delta\mId)^{-1}\mF^H\mF(\mF^H\mF + \delta\mId)^{-1}}_{\mQ, L \times L}\mathcal{F}
\]
which has entries
\[
r(t,s) = \sigma^2\sum_{k,l}Q_{k,l}e^{j(\omega_k t - \omega_l s)}.
\]
The variance will be the trace of this kernel matrix, as
\begin{align*}
    \mathbb{E}[\vert\vert \mathcal{F^*}(\mF^H\mF &+ \delta\mId)^{-1}\mathcal{F}\vn\vert\vert^2_{L_2}] = \int_{0}^{T}r(t,t)dt\\
    &=\sigma^2 \sum_{k,l}Q_{k,l}\int_{0}^{T}e^{j(\omega_k - \omega_l)t}dt\\
    &=2\sigma^2\sum_{k,l}Q_{k,l}e^{j(\omega_k -\omega_l)\frac{T}{2}}\cdot\frac{\sin((\omega_k - \omega_l)\frac{T}{2})}{\omega_k - \omega_l}
\end{align*}
The entries of $\mQ$ can be computed explicitly, providing an estimate for the variance. Figure \ref{fig:variance_vs_snr} presents the variance computed using the preceding expression for various values of \( \gamma \), plotted in multiples of the lower bound defined in \eqref{eq:gamma_lb}. The results indicate that, as the number of snapshots increases, the variance approaches the ideal value \( \frac{\sigma^2}{M} \) (shown in blue), effectively recovering the ideal array gain in the limit as \( N \rightarrow \infty \).

For the bias, we will assume that the signal $\vs$ itself is also a Gaussian random process, generated as
\[
\alpha_n \sim \mathcal{N}(0,\lambda_n),
\]
 where $\lambda_n$ is the $n$th largest eigenvalue of the operator defined in \eqref{eq:operator}. This is equivalent to letting $s(t)$ be a Gaussian random process with a flat power spectral density bandlimited to $\Omega$ that has been observed on the interval $[0,T]$. If we take $\mP = (\mF^H\mF + \delta\mId)^{-1}\mF^H$, then we will have 
 \[
 \mathbb{E}[\vert\vert \text{bias}\vert\vert_2^2] = \text{trace}((\mathcal{F}^*\mP\mathcal{A}-\mathcal{S})\Lambda(\mathcal{A}^*\mP^H\mathcal{F} - \mathcal{S}^*)),
\]
where the $\text{trace}(\cdot)$ of a continuous operator is now the integral along its diagonal (which is equal to the sum of its eigenvalues, see for example \cite{hutson2005ap}). Figure \ref{fig:bias_vs_gamma} illustrates how the bias changes with $\gamma$ for a fixed number of snapshots. The continuous time operators $\mathcal{F}^*$ and $\mathcal{S}^*$ are implemented as matrix representations obtained through dense sampling of rows over the interval $[0, T]$. A detailed account of this construction can be found in \cite{rbip}.

When $\gamma$ is at or near its lower bound (indicated by the dotted red line), the bias is significantly high. Increasing $\gamma$ slightly above the bound given in \eqref{eq:gamma_lb} leads to a significant reduction in bias. For reference, we also show the bias obtained using a Fourier series that is essentially $L$ sinusoids with $\gamma$ set to the lower bound in \eqref{eq:gamma_lb}. As expected, the Fourier series exhibits sublinear convergence with respect to the number of basis functions and eventually plateaus at a relatively high bias level.

\section{Beamspace applications}\label{sec:beam_apps}

The transformation step of our algorithm, as outlined in Section \ref{sec:transform}, maps the array measurements, $\vy$, into a beamspace where the beam vectors are arranged according to their respective directions. Direct access to these spatially separated coefficients enables more efficient computations for tasks that depend on the directional characteristics of the incoming signal. Here, we examine two particular applications of the proposed beamspace transformation: off-grid angle interpolation and interference cancellation.

 \subsection{Off-grid angle interpolation}
The FBST algorithm, as initially formulated, operates on a discrete angular grid, meaning that beamforming is performed only in predefined directions. However, in practical scenarios, signals often arrive from off-grid angles, which do not perfectly align with the sampled directions. This beam misalignment can lead to substantial  performance degradation. There is likely to be reduced array gain and a loss of frequency invariance in the steered direction. Furthermore, accounting for off-grid angles is of particular interest to various processing tasks such as broadband source localization \cite{delude2023it}.

To address this, we extend the algorithm to support arbitrary angles by employing interpolation techniques. Suppose an off-grid angle $\theta^*$ lies between the sampled angles $[\theta_{b},\theta_{b'}]$, and let $\mathcal{I}\equiv\{\vw_{b},\hdots,\vw_{b'}\}$ denote the beam vectors for all the sampled angles within this range. Then we can form an accurate estimate of $\vw^*$ by interpolating with the beam vectors in $\mathcal{I}$. Furthermore, this technique is time-efficient since $|\setI|$ is small compared to the total number of beams.

Figure \ref{fig:super_res} plots a zoomed-in window of the beam pattern in the vicinity of the steered direction $\theta^* = \frac{\pi}{3}$ for a 128-element ULA. The on-grid beams used for interpolation are plotted in blue, and the true beam is plotted in black. Spline interpolation was used for forming the interpolated beam (illustrated in green). Although due to the oversampling factor we form about twice the number of beams, 256 in total, on a 128-element ULA, we just require 8 beams for interpolation to the off-grid angle. Therefore there is negligible additional computational cost while recovering almost the same beamforming performance illustrated in Figure \ref{fig:super_res}. As we will see next, beam interpolation can also be tailored for interference cancellation in the beamspace, allowing off-grid interference suppression with fewer computations.



\subsection{Interference cancellation}\label{sec:interference_cancellation}

\begin{figure*}[t]
    \subfigure[]{
    \includegraphics[width=0.23\linewidth]{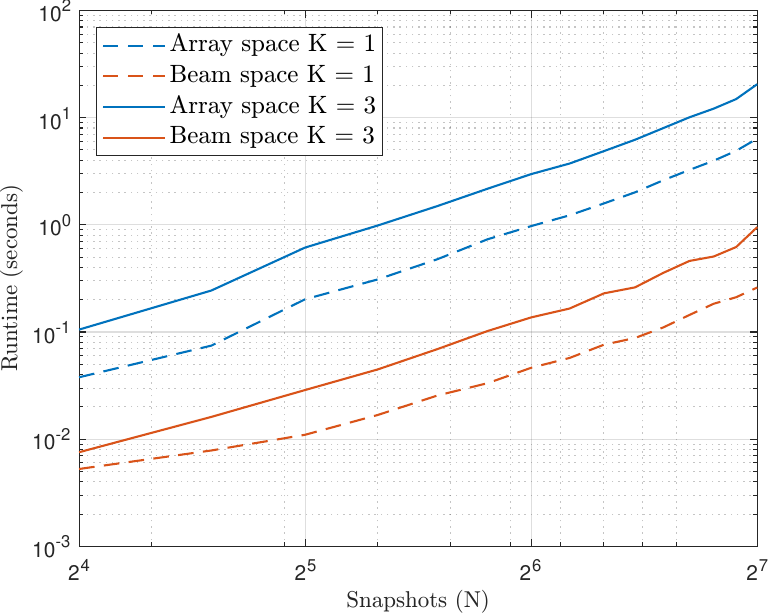}
    \label{fig:interference_runtime}
    }
    \subfigure[]{
    \includegraphics[width=0.23\linewidth]{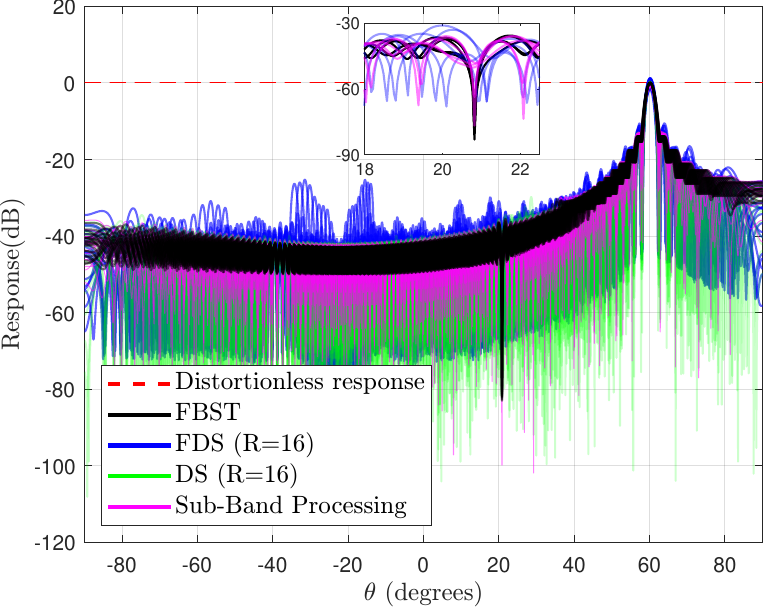}
    \label{fig:beam_pattern_ula}
    }
    \subfigure[]{
    \includegraphics[width=0.23\linewidth]{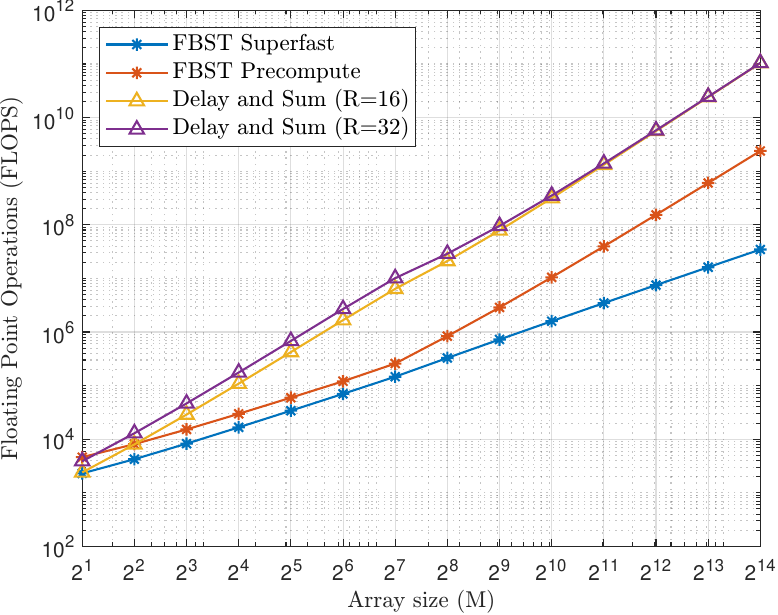}
    \label{fig:ula_runtime_nufft}
    }
    \subfigure[]{
    \includegraphics[width=0.23\linewidth]{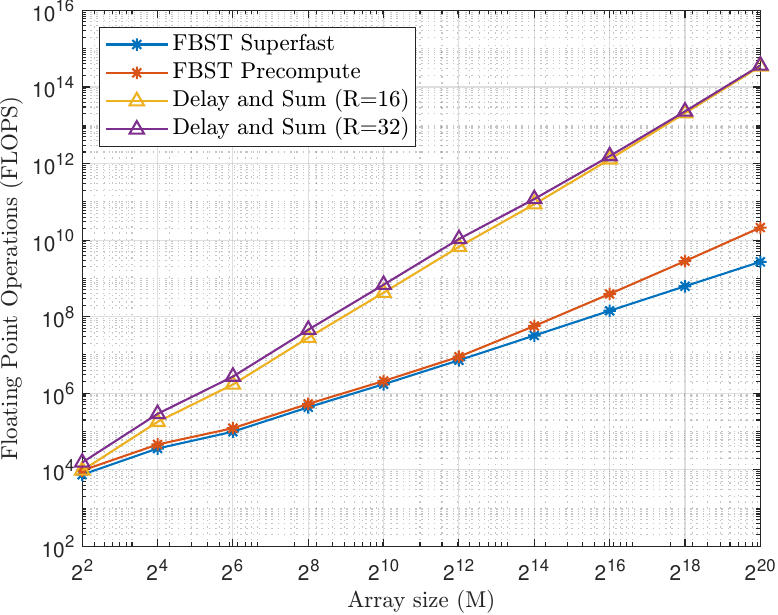}
    \label{fig:upa_runtime_nufft}
    }
    \caption{\sl (a) Runtime for interference cancellation vs. number of snapshots. The experiment was done for single and multi-interference cancellation. In both cases, we observe that the beamspace-based nulling is much faster than computing the projection in array space. (b) Beam pattern for a 128-element ULA after interference cancellation with 64 snapshots. R=16 tap length is used for DS and FDS. Beam patterns for 20 frequencies in $[f_c - \Omega, f_c + \Omega]$ are superimposed to get the frequency response over the entire band. The inlaid plot provides a zoomed-in view of the null placed at the off-grid interfering angle for FBST, FDS, and sub-band processing for a subset of the frequencies. The FDS and sub-band processing nulls are shallower with fluctuations across frequencies. (c) No. of floating point operations vs. array size for a Linear array with non-uniform array spacing. The elements were perturbed with random offsets in their position. The number of beams scales as $B\approx M$ for all algorithms. Two filter lengths $R=16,32$ are used for measuring the DS runtime. (d) No. of floating point operations as a function of array size for a planar array where the sensors are arranged in a square grid of dimensions \( \sqrt{M} \times \sqrt{M} \). The elements were perturbed with random offsets in their position to make the arrangement non-uniform. For all algorithms, the number of beams scales approximately linearly with the number of sensors, i.e., \( B \approx M \). Two filter lengths $R=16,32$, are used for measuring the DS runtime.}
\end{figure*}

Consider the scenario where, in addition to noise, the incoming source signal is also attenuated by an interference signal coming from direction $\theta_I$. Suppose the interference signal is strong and the sidelobes of the beam pattern are not low enough at the interference direction $\theta_I$ (illustrated in Figure \ref{fig:beam_pattern_ula}). In that case, the interference can leak into the main signal and cause distortion. Therefore, to remove this artifact, we use an approximate\footnote{We could easily make this and exact projection. However, the approximated form allows us to directly leverage our fast computational framework.} linear projection that nulls out the interfering signal from the array measurements $\vy$. The array measurements $\vy$ are projected into a subspace orthogonal to the Fourier extension matrix $\mF_I$ corresponding to the interference signal.
\begin{equation*}
    \Tilde{\vy} = \widehat{\mP}^{\perp}_I\vy = (\mId - \widehat{\mP}_I)\vy = (\mId - \mF_I(\mF_I^H\mF_I + \delta\mId)^{-1}\mF_I^H)\vy.
\end{equation*}
The FBST algorithm can now be applied to the modified array measurements $\Tilde{\vy}$.

While projection-based interference removal effectively eliminates unwanted signals, it also inherently modifies the signal subspace. Since $\widehat{\mP}^{\perp}_{I}$ removes any component aligned with the interferer subspace, it alters the structure of the desired signal, leading to potential performance degradation. This interference bias can be formally quantified using the functions and notations introduced in Section \ref{sec:err}. 

\begin{figure}
     \centering
     \includegraphics[width=0.32\linewidth]{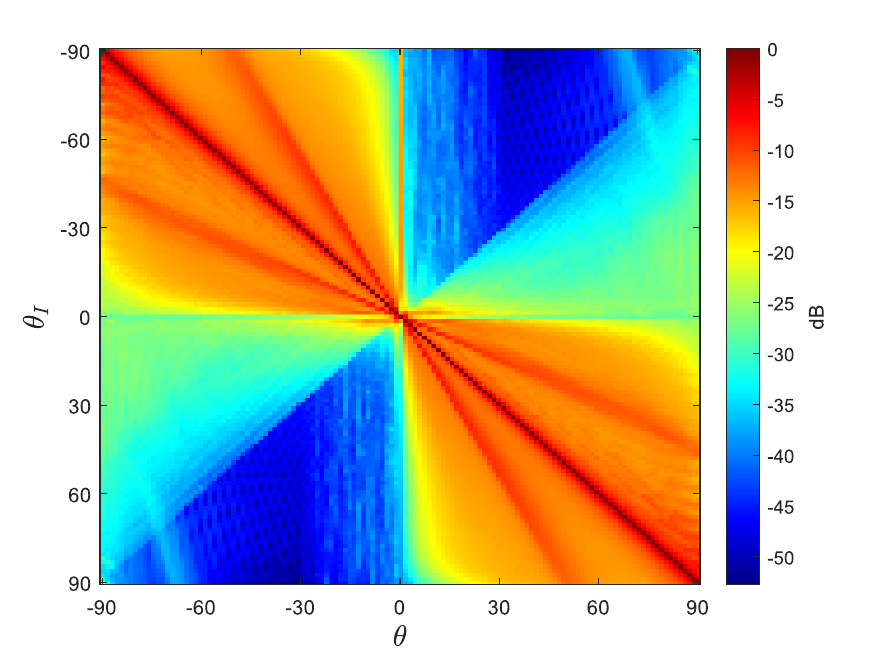}
     \caption{\sl Interference bias for a 128-element ULA with 64 snapshots. The interference bias is small when the source and interference are spatially separated; as they come closer, a larger part of the source signal subspace bleeds into the interferer subspace, causing the interference nulling to remove a significant portion of the source signal.}
     \label{fig:interference_bias}
 \end{figure}

Let $\valpha$ and $\valpha_I$ denote the signal and interference Slepian coefficients respectively. We can then define the error vector
\[
\ve_I = \mathcal{F}\hat{\vbeta} - \mathcal{F}\tilde{\vbeta},
\]
where $\hat{\vbeta} = (\mF^H\mF + \delta\mId)^{-1}\mF^H(\mathcal{A}\valpha)$ and $\tilde{\vbeta} = (\mF^H\mF + \delta\mId)^{-1}\mF^H\widehat{\mP}^{\perp}_I(\mathcal{A}\valpha + \mathcal{A}_I\valpha_I)$ are the least squares estimate for Fourier coefficients of an uncorrupted source signal and an interference contaminated source signal respectively. Assuming that both the source and interfering signals are independent Gaussian processes, it is easy to show that the interference bias $\mathbb{E}[\vert\vert \ve_I\vert\vert^2_2]$ takes the form
\begin{align*}
    \mathbb{E}[\vert\vert \ve_I\vert\vert_2^2] &= \text{trace}((\mathcal{F}^*\mP\mP_I\mathcal{A})\Lambda(\mathcal{A}^*\mP_I^H\mP^H\mathcal{F}))\\
    &+ \text{trace}((\mathcal{F}^*\mP\mP^{\perp}_I\mathcal{A}_I)\Lambda_I(\mathcal{A}_I^*(\mP_I^{\perp})^H\mP^H\mathcal{F})).
\end{align*}
Figure \ref{fig:interference_bias} plots the expression given above. We can observe that the interference bias remains relatively small but increases as the interfering signal's direction approaches that of the source signal. This occurs because a larger portion of the source subspace overlaps with the interferer subspace, leading to greater distortion.

 As mentioned before, interference nulling can also be done directly in the beamspace. The Fourier coefficients estimate after interference cancellation is given by $\tilde{\vbeta} = (\mF^H\mF + \delta \mId)^{-1}\mF^{H}\Tilde{\vy}$. It follows from our discussion in Section \ref{sec:form} that this is mathematically equivalent to applying FBST on $\Tilde{\vy}$. Next, we expand $\Tilde{\vy}$ to get,
 \[
\tilde{\vbeta} = (\mF^H\mF + \delta \mId)^{-1}\mF^{H}(\vy - \mF_I(\mF_I^H\mF_I + \delta\mId)^{-1}\mF_I^H\vy).
 \]
Let $\vw = \mF^H\vy$ and $\vw_I = \mF_I^H\vy$, then the above equation can be reformulated in beamspace as
\[
\tilde{\vbeta} = (\mF^H\mF + \delta \mId)^{-1}(\underbrace{\vw}_{\mF^H\vy} - \underbrace{\mF^H\mF_I(\mF_I^H\mF_I + \delta\mId)^{-1}}_{{\mG}}\underbrace{\vw_I}_{\mF_I^H\vy}).\]
It is also straightforward to see that for $P$ interferers, the beamspace formulation generalizes to,
\begin{equation}\label{eq:interfer_beam_space}
    \tilde{\vbeta} = (\mF^H\mF + \delta \mId)^{-1}\left(\vw - \sum_{p=1}^{P}\mG_p\vw_{p}\right).
\end{equation}
Interference cancellation in beamspace is significantly faster than performing it directly in array space (i.e., using $\vy$). Due to the directional separation of the beams, we only need to work with the $L$ entries of the beam vectors rather than processing all $MN$ measurements.

The preceding nulling approach manipulates beamspace vectors in the frequency domain, but an equivalent time-domain representation can be obtained by properly formulating the signal reconstruction process. The fan-like transform applied to the array measurements yields beams $\{\vw_b\}_{b=1}^B$, which directly correspond to Fourier-domain weight vectors $\{\boldsymbol{\beta}_b\}_{b=1}^B$ via the relation
\[
\vbeta_b = (\mF_b^H\mF_b + \delta \mId)^{-1} \vw_b,
\]
where $\mF_b$ denotes the basis matrix for the $b^{\text{th}}$ angle. With a reconstruction matrix $\mF_u$, we extend this mapping to the time domain, producing time-domain beams $\{ \vy_b \}_{b=1}^{B}$ given by $\vy_b = \mF_u \vbeta_b$. 

The vector $\tilde\vbeta$ defined in \eqref{eq:interfer_beam_space} can be expressed as a combination of the signal weights $\boldsymbol{\beta} = (\mF^H\mF + \delta \mId)^{-1} \vw$ and interference weights ${\vbeta}_p$ as
\[
\tilde{\vbeta} = \boldsymbol{\beta} - \sum_{p=1}^p \mH_p {\vbeta}_{p},
\]
with $\mH_p = (\mF^H\mF + \delta \mId)^{-1} \mF^H \mF_p$. The corresponding time-domain beam $\tilde{\vy}_s$ becomes
\[
\tilde{\vy}_s = \mF_u \boldsymbol{\beta} - \mF_u \sum_{p=1}^P \mathbf{H}_{p} \boldsymbol{\beta}_{p}.
\]
If the matrix $\mF_u$ is defined over a Nyquist interval, it tends to be fat. However, with slight oversampling and a broader reconstruction interval $T$, it can approach full column rank, allowing $\mF_u^{\dagger} \mF_u \approx \mathbf{I}$. This leads to the reformulation:
\begin{align}\label{eq:time_beam}
    \tilde{\vy}_s &= \underbrace{\mF_u \vbeta}_{\vy_s} - \sum_{p=1}^P \underbrace{\mF_u \mH_{p} \mF_u^{\dagger}}_{\mG'_{p}} \mF_u \boldsymbol{\beta}_{p} = \vy_s - \sum_{p=1}^P \mG'_{p} \vy_{p}.
\end{align}
This interference-suppressed time-domain beam can then be windowed over the Nyquist interval or interpolated to other regions of interest. As this formulation operates in the time domain, it is naturally extensible to DS and FDS methods that directly output multiple beams in the time-domain. 

It should be noted that the final expressions in \eqref{eq:interfer_beam_space}, \eqref{eq:time_beam} depend only on the beam vectors associated with the directions we intend to suppress, regardless of whether the underlying representation is in the time domain or the frequency domain.  As a result, exact alignment of the interference directions with the angular grid is not required. High-resolution estimates of off-grid beam vectors can be obtained using the interpolation approach introduced in the previous section. Figure \ref{fig:interference_runtime} illustrates the runtime comparison of interference nulling for off-grid interferers when performed in beamspace versus array space. The results indicate that interference suppression in beamspace, both for single and multiple interferers, is computationally more efficient. Additionally, the interpolation used in beamspace for off-grid interferers does not introduce a significant computational overhead, as it involves only a small number of beam vectors.

Lastly, Figure \ref{fig:beam_pattern_ula} plots the beam patterns for FBST, DS, FDS and sub-band processing on a 128-element ULA steered at $\theta \approx 60^{\circ}$ with a manual null placed at an off-grid interference angle $\theta_I \approx 20^{\circ}$. In the case of FBST, the beam vector corresponding to the interference direction was estimated using a collection of 8 surrounding on-grid beam vectors. A similar approach was used for sub-band processing with interpolation done for each beam-coefficient for all the frequency bins. As shown, the null produced by FBST is noticeably deeper and remains stable across the entire frequency band, in contrast to the DS, FDS and sub-band processing methods, which exhibit greater variability due to their inherent bias and low frequency resolution. In addition to the observed variations in the nulls, the artifacts appearing in the steered directions, as discussed in Fig. \ref{fig:ula_beampattern}, persist for FDS and sub-band processing in the interference nulling scenario.

\section{Non-uniform Array spacing}\label{sec:nufft_1}

Until now, we have considered linear and planar arrays where sensors were uniformly placed however it is straightforward to generalize our algorithm to irregular array positions. We begin by revising the summation in \eqref{eq:DFT_1} to write,
\begin{align}\label{eq:DFT_non_unif}
    w_b(\ell) = \left[\mF_{\theta_b}^\H\vy\right]_\ell &= \sum_{m,n}y_{m,n}e^{j\pi(\zeta + \xi \ell' )b' \mathcal{G}(m)}e^{-j\frac{2\pi\epsilon}{L}\ell'n},
\end{align}
where $\mathcal{G}(\cdot)$ is a function that maps the $m^{th}$ sensor to its position. The summation over $n$ remains unaffected and can be applied again using a chirp $z-$transform. When $\mathcal{G}$ is affine, the summation over $m$ can be modified to resemble a chirp $z-$transform and applied at no additional cost. For any arbitrary $\mathcal{G}$ the summation over $m$ is a classical example of a type-I Non-uniform FFT (NUFFT)\cite{nufft1,nufft2,nufft3} and can be again applied in log-linear time with an additional computational overhead of $\mathcal{O}(\log \frac{1}{\Delta}M)$ where $\Delta$ is the required level of accuracy. The rest of the FBST framework and the beamspace applications remain unaltered since the irregular spacing only affects the summation in \eqref{eq:DFT_non_unif} that is particular to the Fan FFT step.

Figures \ref{fig:ula_runtime_nufft}, \ref{fig:upa_runtime_nufft} plot the total floating point operations (FLOPS) per-sample when using NUFFT. Non-uniform array spacing was emulated by adding random perturbations to the ULA, UPA experiments. The FBST is significantly more scalable than delay and sum for non-uniform array spacing, just like in the uniform scenario. The extra calculations needed for NUFFT don't substantially increase the algorithm's processing burden. We omit the beamforming performance comparison here since $\Delta$ can be tuned to several orders below the observed bias, providing performance gains indistinguishable from the uniform case.
 
\bibliographystyle{IEEEtran}
\bibliography{IEEEabrv.bib,ref.bib}


\begin{appendices}\label{upa_f}

    \section{FDS for planar array}
    Suppose that we have a planar array in the $(x,y)$ plane with $M$ elements located at $\{(x_m, y_m)\}$. The angle of elevation $\theta$ is measured from the $z$ axis, and the azimuthal angle $\phi$ is measured from the $x$ axis to the projection of the arriving signal. If there is a signal incident from $(\theta,\phi)$ which we call $s(t)$ at the origin, then the signal at element $m$ is,
    \[
    s_m(t) = s(t - \tau_m),
    \tau_m = \frac{x_m\sin\theta\cos\phi + y_m\sin\theta\sin\phi}{c}.
    \]
    When performing delay and sum computations, we want to evaluate,
    \[
    y_{\theta,\phi}(t) = \sum_m s_m(t + \frac{x_m\sin\theta\cos\phi}{c} + \frac{y_m\sin\theta\sin\phi}{c}).
    \]

    Assuming that the spacing in the planar array is uniform and at half wavelength, the array locations can be written as a separable set $(x_m,y_m) \in \{x_1,\hdots,x_{\sqrt{M}}\}\otimes\{y_1,\hdots,y_{\sqrt{M}}\}$ and the above sum can be decomposed as,
    \[
    y_{\theta,\phi}(t) = \sum_{m_1}\sum_{m_2}s_{m_1,m_2}(t + \frac{x_{m_1}\sin\theta\cos\phi}{c} + \frac{y_{m_2}\sin\theta\sin\phi}{c}).
    \]
    Now, to evaluate the above summation using Fast delay and sum, we introduce separability in the beam directions. Specifically, we set,
    \[
    \begin{bmatrix}
    \mu\\
    \kappa
\end{bmatrix} = \begin{bmatrix}
    \sin\theta\cos\phi\\
    \sin\theta\sin\phi
\end{bmatrix}\iff \begin{bmatrix}
    \phi\\
    \theta
\end{bmatrix} = \begin{bmatrix}
    \tan^{-1}(\frac{\kappa}{\mu})\\
    \sin^{-1}(\sqrt{\kappa^2 + \mu^2})
\end{bmatrix}
    \]
    and sample $(\mu,\kappa)$ such that,
    \begin{align*}
        \mu_{r_1} &= 1 - \frac{2r_1 + 1}{\sqrt{M}} \ r_1\in \{0,1,\hdots,\sqrt{M}-1\}\\
        \kappa_{r_2}&= 1 - \frac{2r_2 + 1}{\sqrt{M}} \ r_2\in \{0,1,\hdots,\sqrt{M}-1\}.\\
    \end{align*}
    Then, the delay and sum expression can be rewritten as,
    \[
    y_{r_1,r_2}(t) = \sum_{m_1}\sum_{m_2}s_{m_1,m_2}(t + \frac{x_{m_1}\mu_{r_1}}{c} + \frac{y_{m_2}\kappa_{r_2}}{c})
    \]
    and we can compute
    \[
    v_{m_1,r_2}(t) = \sum_{m_2}s_{m_1,m_2}(t + \frac{y_{m_2}\kappa_{r_2}}{c}).
    \]
    Due to the sampling over $\kappa$ the above summation can be transformed into a radix-2 beamformer \cite{pruned} and evaluated using fast delay and sum in $\mathcal{O}(\sqrt{M}\log\sqrt{M})$ operations for each $m_1$, requiring a total of $\mathcal{O}(M\log\sqrt{M})$ operations.
    To get the final beamformed signal, we write,
    \[
    y_{r_1,r_2}(t) = \sum_{m_1}v_{m_1,r_2}(t + \frac{x_{m_1}\mu_{r_1}}{c}).
    \]
    The above summation can again be implemented using FDS, requiring a total of $\mathcal{O}(M\log\sqrt{M})$ operations. Therefore, to form $M$ beams we need $\mathcal{O}(M\log\sqrt{M})$ operations per sample.

    \section{FBST for planar array}\label{appdx:fbst_upa}
    Consider an $M$-element planar array with uniformly spaced sensors arranged in a $\sqrt{M}\times\sqrt{M}$ rectangular grid. A broadband signal $s(t)$ impinges the array at a spherical angle $\theta = (\varphi,\phi)$. The time delays across the planar array can be written as,
    \begin{equation*}
        \tau_{k,m}(\varphi,\phi) = \frac{k}{2\tilde{f}_c}\sin\varphi\sin\phi + \frac{m}{2\tilde{f}_c}\sin\varphi\cos\phi,
    \end{equation*}
    where $k,m$ are the array indexing and $\tilde{f}_c = f_c + \Omega$. Correspondingly the $l^{th}$ entry of $\mF_{\theta}^H\vy$ in \eqref{eq:betahatls} can be written as,
    \begin{align*}
        \left[ \mF^{H}_{\theta}\vy \right]_{\ell} &= \sum_{n=1}^{N}\sum_{k=1}^{\sqrt{M}}\sum_{m=1}^{\sqrt{M}}y_{k,m}[n]e^{j(2\pi (f_c+\omega_{\ell})\tau_{k,m}(\varphi,\phi)-\omega_{\ell}t_n)}\\
        &= \sum_{n=1}^{N}\sum_{k=1}^{\sqrt{M}}\sum_{m=1}^{\sqrt{M}}y_{k,m}[n]e^{-j\omega_1(\ell)k}e^{-j\omega_2(\ell)m}e^{-j\omega_3(\ell)n}
    \end{align*}
    where 
    \begin{align*}
        \omega_1(\ell) &= -\frac{\pi \sin\varphi\sin\phi}{\tilde{f}_c}\left(f_c + \frac{\epsilon}{LT_s}\cdot(\ell - \frac{L}{2})\right)\\
        \omega_2(\ell) &= -\frac{\pi\sin\varphi\cos\phi}{\tilde{f}_c}\left(f_c + \frac{\epsilon}{LT_s}\cdot(l - \frac{L}{2})\right)\\
        \omega_3(\ell) &= \frac{2\pi\epsilon}{L}\cdot\left(\ell - \frac{L}{2}\right)
    \end{align*}

\begin{figure}
        \centering
        \includegraphics[width=0.55\linewidth]{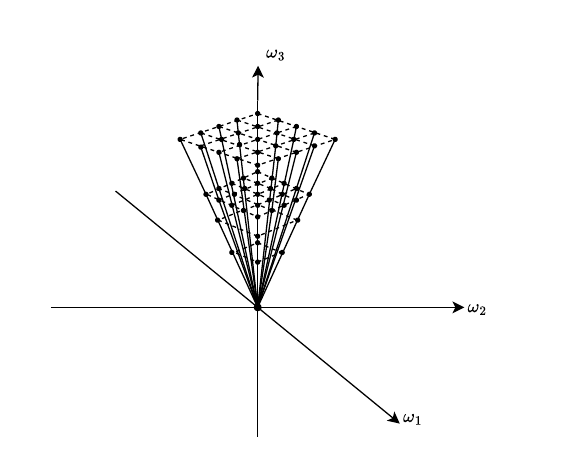}
        \caption{\sl A pseudo polar sampling grid in the 3D Fourier domain. The samples are equispaced along the $\omega_3$ axis and in slope along each slanted line.}
        \label{fig:polar_3D}
    \end{figure}

    Analogous to the linear case, this basically means that $\mF_{\theta}^H\vy$ is a collection of $L$ equally spaced samples along a slanted line of the 3D discrete time Fourier transform of $\{y_{k,m}[n]\}_{k,m,n}$. Same as before, we get a temporal frequency $\omega_{3}(\ell)$ for fluctuations in the signal across time. However, compared to the linear case, in the planar array we get two spatial frequencies $\omega_1(\ell)$ and $\omega_2(\ell)$, one for each array axis. When beamforming in multiple directions $(\varphi,\phi)$, we get multiple slanting lines in the 3D frequency domain as illustrated in Figure \ref{fig:polar_3D}. Next, suppose we are interested in forming $B$ beams in total with $\sqrt{B}$ beams in the elevation and azimuth coordinates, respectively, where, for simplicity, we assume that $B$ is a perfect square. To mold the above formulation into the Fan FFT step given in section \ref{sec:transform} we use the separability trick introduced in the previous appendix. We basically discretize $\mu$ and $\kappa$ from the previous appendix using indices $(a,b)$ to get a sampling similar to \eqref{eq:thetab} that is equispaced in slope in the 3D frequency domain.
    This allows us to rewrite the expression for $\mF_{\theta}^H\vy$ as,
    \begin{align*}
        w_{a,b}(\ell) &= \left[\mF_{\theta_{a,b}}^H\vy\right] = \sum_m e^{j \pi(\zeta + \xi\ell^{'}) b'm}\\
        &\cdot \sum_k e^{j \pi(\zeta + \xi\ell^{'}) a' k}\sum_n y_{k,m}[n] e^{-j \frac{2\pi\epsilon}{L}\ell^{'} n},
    \end{align*}
    where $\zeta = \frac{2f_c}{(\sqrt{B}-1)\tilde{f}_c}$, $\xi = \frac{2\epsilon}{(\sqrt{B}-1)LT_s\tilde{f}_c}$, $a' = a - \frac{\sqrt{B}+1}{2}$, $b' = b - \frac{\sqrt{B}+1}{2}$ and $\ell^{'} = \ell - \frac{L}{2}$. Now the innermost summation in the above expression can be written as a chirp $z$ mapping $N$ time samples to $L$ frequency samples for each ordered pair $(k,m)$. Thus, the number of operations required is $\mathcal{O}(MN\log N)$. The summation over $k$ is again a chirp $z$, this time mapping $\sqrt{M}$ array samples to $\sqrt{B}$ frequency samples for each ordered pair $(m,\ell)$. Hence, the number of operations in this case is $\mathcal{O}(\sqrt{M}N\sqrt{B}\log\sqrt{B})$. Finally, the summation over $m$ is also a chirp $z$ again mapping $\sqrt{M}$ array samples to $\sqrt{B}$ frequency samples for each ordered pair $(a,\ell)$. Therefore, the cost of this summation is $\mathcal{O}(\sqrt{M}N\sqrt{B}\log\sqrt{B})$.

    The total cost of applying FanFFT projection to a planar array therefore, amounts to $\mathcal{O}(MN\log N + \sqrt{M}\sqrt{B}N\log \sqrt{B})$ operations or $\mathcal{O}(M\log N + \sqrt{M}\sqrt{B}\log \sqrt{B})$ operations per sample. When forming $B\approx M$ beams, this reduces to $\mathcal{O}(M\log N + B\log\sqrt{B})$ i.e., similar to the linear array case. Since the inversion step and the beamspace applications are independent of the underlying array geometry, their application to the UPA remains unchanged, provided that the adaptations for the 3D Fan-FFT are accounted for.
    It is important to note that the separability based sampling procedure may generate a few invalid beam directions, since the $\arcsin$ function is only defined over the interval $[-1,1]$. Nevertheless, such vacuous beams are limited in number, and the total valid beam count still scales as $\mathcal{O}(B)$.

\end{appendices}


\vfill

\end{document}